\documentclass[a4paper,10pt]{article}
\usepackage{subfigure}
\usepackage{amsfonts}
\usepackage[dvips]{color,graphicx}
\usepackage{graphicx,amssymb,amsmath,bm,latexsym}
\usepackage{stmaryrd}
\usepackage{bm}
\usepackage{CJK}
\usepackage{indentfirst}
\usepackage{amsmath}

\definecolor{red  }{rgb}{1,0,0}
\definecolor{blue }{rgb}{0,0,1}
\definecolor{green}{rgb}{0,1,0}

\DeclareMathOperator{\sech}{sech}

\usepackage{graphicx,amssymb,amsmath,bm,latexsym}

\begin{document}
\vspace{-2.3em}
\title{A novel (2+1)-dimensional integrable KdV equation with peculiar solution structures}
\author{S. Y. Lou\thanks{E-mail:lousenyue@nbu.edu.cn.}\\
\footnotesize \it School of Physical Science and Technology, Ningbo University, Ningbo, 315211, China}
\date{\today}
\maketitle

\vspace{-1.5em}
\begin{abstract}
The celebrated (1+1)-dimensional Korteweg de-Vries (KdV) equation and its (2+1)-dimensional extention,
the Kadomtsev-Petviashvili (KP) equation, are two of the most important models in physical science. 
The KP hierarchy is explicitly written out by means of the linearized operator of the KP equation. A novel (2+1)-dimensional KdV extension, the cKP3-4 equation, is obtained by combining the third member (KP3, the usual KP equation) and the fourth member (KP4) of the KP hierarchy. The integrability of the cKP3-4 equation is guaranteed by the existence of the Lax pair and dual Lax pair. The cKP3-4 system can be bilinearized by using Hirota's bilinear operators after introducing an additional auxiliary variable. Exact solutions of the cKP3-4 equation possess some peculiar and interesting properties which are not valid for the KP3 and KP4 equations. For instance, the soliton molecules and the missing D'Alembert type solutions (the arbitrary travelling waves moving in one direction with a fixed model dependent velocity) including periodic kink molecules, periodic kink-antikink molecules, few cycle solitons and envelope solitons are existed for the cKP3-4 equation but not for the separated KP3 equation and the KP4 equation.
\end{abstract}

\section{Introduction \vspace{-0.3em}}
In linear science, the wave motion equation 
\begin{equation}
u_{tt}-c^2u_{xx}=0\label{WME}
\end{equation}
is used to describe almost all the waves in natural science because the general solution, the D'Alembert solution, 
\begin{equation}
u=f(x-|c|t)+g(x+|c|t), \label{ALBE}
\end{equation}
includes arbitrary waves ($f$ and $g$ are arbitrary functions of the indicated variables) moving in the $x$ direction or in the $-x$ direction with a fixed velocity $|c|$.

The simplest extension of the wave motion equation \eqref{WME} in nonlinear science is the so-called Korteweg de-Vries (KdV) equation \cite{KdV}
\begin{equation}
u_t + u_{xxx} + 6uu_x = 0. \label{KdV}
\end{equation}
 The KdV equation
possesses various applications in physics and other scientific fields. It approximately describes the evolution of long, one-dimensional waves in many physical settings, including shallow-water waves with weakly non-linear restoring forces, long internal waves in a density-stratified ocean, ion acoustic waves in a plasma, acoustic waves on a crystal lattice \cite{KdV} and the 2-dimensional quantum gravity \cite{Guo}. The KdV equation can be solved by using the inverse
scattering transform \cite{IST} and other powerful methods such as the Hirota's bilinear method \cite{Hirota} and the Darboux transformation \cite{Gu}.

The KdV equation and other nonlinear extensions lost the D'Alembert type solutions (arbitrary waves moving in one direction with a fixed model dependent (material dependent) velocity. Thus, one can naturally ask the following important question:

\em Can we find a nonlinear extension such that the missing D'Alembert type waves can be found? \rm

To describe two dimensional nonlinear waves which are the extensions of the (1+1)-dimensional KdV waves, there are some different versions of the (2+1)-dimensional integrable KdV equations including the Kadomtsev-Petviashvili (KP, or KP3) equation \cite{KP}
\begin{equation}
 u_t-a\left(6uu_x+u_{xxx}-3w_{y}\right)=0,\ u_y=w_x, \label{KP}
\end{equation}
the Nizhnik-Novikov-Veselov (NNV)
equation\cite{NNV1,NNV2,NNV3}
\begin{equation}
u_t+u_{xxx}+u_{yyy}+3\left(uv\right)_x+3\left(uw\right)_y=0,\ u_x=v_y,\ u_y=w_x, \label{NNV}
\end{equation}
the asymmetric NNV equation (ANNV) \cite{ANNV1,ANNV2}, or named Boiti-Leon-Manna-Pempinelli equation\cite{ANNV3}
\begin{equation}
u_t+u_{xxx}+3\left(uv\right)_x=0,\ u_x=v_y, \label{ANNV}
\end{equation}
the Ito equation \cite{Ito}
\begin{equation}
au_t+u_y+u_{xxx}+3uu_x+3v_x+bu_x=0,\ u_y=w_x,\ v_y=wu_x, \label{Ito}
\end{equation}
and the breaking soliton equation \cite{BS,BS1}
\begin{equation}
u_t+u_{xxy}+4uw_x+2wu_x=0,\ u_y=w_x. \label{BS}
\end{equation}
It is clear that when $y=x$ all the (2+1)-dimensional integrable KdV equations \eqref{KP}--\eqref{BS} are reduced back to the equivalent KdV equation \eqref{KdV}.

The stability and dynamics of soliton molecules (soliton bound states) have attracted considerable attention in several areas such as the optics \cite{SP05,HK17,LXM,Nano,SA} and  Bose-Einstein condensates \cite{LNS}.  Some theoretical proposals to form soliton molecules have been established \cite{CK03,YB11}. Especially, in Refs. \cite{LouS1,LouS2}, a new mechanism, the velocity resonance, to form soliton molecules is proposed. It is found that the standard (1+1)-dimensional KdV equation \eqref{KdV} does not possess soliton molecules. However, in real physics higher order effects such as the higher order dispersions and higher order nonlinearities have been neglected when the KdV equation \eqref{KdV} is derived \cite{KdV5}. Whence the higher order effects are considered to the usual KdV equation, one can really find the soliton molecules \cite{LouS1}. By using the velocity resonance  mechanism, some authors find new types of soliton molecules such as the dromion molecules and half periodic kink molecules in some physical systems \cite{LiB,XYT,YZW}.

After detailed calculations, one can find that there is no soliton molecules for all the (2+1)-dimensional KdV extensions \eqref{KP}--\eqref{BS}. To find soliton molecules for (2+1)-dimensional KdV systems, we have to consider their higher order effects. For integrable systems, we can directly add the integrable higher order systems to the lower order ones. In section 2 of this paper, we write down the positive KP hierarchy in terms of the linearized operator of the KP equation \eqref{KP}. Then we write down a novel (2+1)-dimensional integrable KdV system, the combination of the KP3 and the KP4 equations (cKP3-4), which will be reduced back to the usual (1+1)-dimensional KdV equation \eqref{KdV} when $y=x$. The Lax pair and the dual Lax pair of the cKP3-4 equation are also given in section 2. In section 3, we study the multiple soliton solutions of the novel (2+1)-dimensional KdV equation. The soliton molecules are investigated in section 4. It is found that a soliton molecule may include arbitrary number of solitons. For this model, a single soliton molecule without other solitons may possess arbitrary shape. This factor can be simply proved by studying its travelling waves. The last section is a summary and discussion.

\section{A novel (2+1)-dimensional KdV extension: cKP3-4 equation}
For a (1+1)-dimensional integrable system, there is a so-called recursion operator $\Phi$ such that a set of infinitely many commute symmetries can be directly obtained $K_n=\Phi^nu_x$. Thus, an integrable hierarchy,
\begin{equation}
u_{t_n}=\Phi^n u_x,\ n=0,\ 1,\ \ldots,\ \infty, \label{1+1}
\end{equation}
can be simply obtained.

However, in (2+1)-dimensional case, there is no such kind of recursion operators to find infinitely many commute symmetries and then the integrable hierarchy. Fortunately, by means of the formal series symmetry approach \cite{PRLou}, for many kinds of (2+1)-dimensional integrable systems including the KP equation \eqref{KP} \cite{JPA93}, the NNV equation \eqref{NNV}\cite{JMP94} and the ANNV equation \eqref{ANNV}, one can find infinitely many commute symmetries by using their linearized operators. For instance, for the KP equation \eqref{KP} with $a=1$, a symmetry, $\sigma$, is defined as a solution of its linearized equation
\begin{equation}
6\sigma u_x+6u\sigma_x+\sigma_{xxx}-3\partial_x^{-1}\sigma_{yy}-\sigma_t\equiv L\sigma=0, \label{KPs}
\end{equation}
where $L\equiv \partial_x^3+6\partial_x u-3\partial_x^{-1}\partial_{y}^2-\partial_t$ is the linearized operator of the KP equation \eqref{KP} and $\partial_x^{-1}$ is defined as $\partial_x^{-1}\partial_x=\partial_x\partial_x^{-1}=1$. In Ref. \cite{JPA93}, it is proven that the following expressions
\begin{equation}
\sigma_n=\frac{1}{2(n-1)!3^n}L^{n}y^{n-1}, n=1,\ 2,\ \ldots \infty, \label{SymKP}
\end{equation}
are all commute symmetries of the KP equation \eqref{KP} for arbitrary integers $n\geq 1$.

Thus we obtain the integrable positive KP hierarchy in the form
\begin{eqnarray}
u_{t_n}=\frac{1}{2(n-1)!3^n}L^{n}y^{n-1}, n=1,\ 2,\ \ldots \infty. \label{2+1}
\end{eqnarray}
The first five of \eqref{2+1} read ($u_y=w_x,\ u_{yy}=z_{xx},\ u_{yyy}=m_{xxx}$),
\begin{eqnarray}
&&\mbox{\rm KP1:}\ u_{t_1}=u_x, \label{KP1}\\
&&
\mbox{\rm KP2:}\ u_{t_2}=-2 u_y, \label{KP2}\\
&&
\mbox{\rm KP3:}\ u_{t_3}=-6uu_x-u_{xxx}+3w_{y},\ \label{KP3}\\
&&
\mbox{\rm KP4:}\ u_{t_4}=12\left(2wu_x-z_y+u_{xxy}+4uu_y\right), \label{KP4}\\
&&
\mbox{\rm KP5:}\ u_{t_5}=u_{xxxxx}+10[(u^3+uu_{xx}-zu-w^2-z_{xx})_x+u_xu_{xx}-uz_x
-\partial_x^{-1}(uz_{xx}+w_x^2)]+5m_y.\ \label{KP5}
\end{eqnarray}
It is not difficult to find that the KP3 equation \eqref{KP3} is just the KP equation \eqref{KP} by taking $t_3=-at$. It is interesting that in addition to the KP3 equation, the KP4 equation is also an extension of the KdV equation. In other words, when $y=x$ the KP4 equation \eqref{KP4} is also reduced back to an equivalent form of the usual (1+1)-dimensional KdV equation \eqref{KdV}. Because the commute KP3 and KP4 equations are all the (2+1)-dimensional KdV extensions, we can obtain a novel integrable (2+1)-dimensional KdV equation, the cKP3-4 equation
\begin{eqnarray}
&&u_{t}=a(uu_x+u_{xxx}-3w_{y})+b\left(2wu_x-z_y+u_{xxy}+4uu_y\right), \label{KP34}\\
&&u_y=w_x,\quad u_{yy}=z_{xx}. \nonumber
\end{eqnarray}
It is trivial that when $y\rightarrow x$,\ $w\rightarrow u,\ z\rightarrow u$, the (2+1)-dimensional KdV equation \eqref{KP34} is reduced back to the usual KdV equation \eqref{KdV} up to some suitable scaling and Galileo transformations.

To see the integrability of the model \eqref{KP34}, we directly write down its Lax pair
\begin{eqnarray}
\psi_{y}&=&\mbox{\rm i}(\psi_{xx}+u\psi),\ \mbox{\rm i}\equiv \sqrt{-1}, \label{psiy}\\
\psi_{t}&=&2\mbox{\rm i}b\psi_{xxxx}+4a\psi_{xxx}+4\mbox{\rm i}bu\psi_{xx}+2(3au+2\mbox{\rm i}bu_x
+bw)\psi_x\nonumber\\
&&-\mbox{\rm i}(3aw+bz-2bu^2+3a\mbox{\rm i}u_x-2bu_{xx}+\mbox{\rm i}bw_x)\psi,\label{psit}
\end{eqnarray}
and the dual Lax pair
\begin{eqnarray}
\phi_{y}&=&-\mbox{\rm i}(\phi_{xx}+u\phi),\  \label{phiy}\\
\phi_{t}&=&-2\mbox{\rm i}b\phi_{xxxx}+4a\phi_{xxx}-4\mbox{\rm i}bu\phi_{xx}+2(3au-2\mbox{\rm i}bu_x
+bw)\phi_x\nonumber\\
&&+\mbox{\rm i}(3aw+bz-2bu^2-3a\mbox{\rm i}u_x-2bu_{xx}-\mbox{\rm i}bw_x)\phi.\label{phit}
\end{eqnarray}
One can directly verified that the compatibility condition
$\psi_{yt}=\psi_{ty}$ (and/or $\phi_{yt}=\phi_{ty}$) is nothing but the field $u$ is a solution of the cKP3-4 equation \eqref{KP34}.

Similar to the KP equation \eqref{KP}, one can also directly verify that
\begin{equation}
u_{t_{-1}}=(\psi\phi)_x, \label{NKP}
\end{equation}
where spectrum functions $\psi$ and $\phi$ defined by \eqref{psiy}--\eqref{phit}, is a negative flow of the cKP3-4 equation. By using the method proposed in Refs. \cite{NKP1,NKP2}, the whole negative KP hierarchy can be obtained from \eqref{NKP} in terms of the Lax operators defined in \eqref{psiy} and \eqref{phiy}.

The existence of the Lax pairs allows one to study the cKP3-4 equation by means of the standard methods such as the inverse scattering transformation, Darboux transformation, $\bar{\partial}$ approach and so on. To study the multiple soliton solutions, the Hirota's bilinear method is a simplest way.

\section{Bilinearization of the cKP3-4 equation \eqref{KP34}}
By making the transformation
\begin{equation}
u=2(\ln f)_{xx},\ w=2(\ln f)_{xy},\ z=2(\ln f)_{yy}, \label{ru}
\end{equation}
the cKP3-4 equation \eqref{KP34} becomes
\begin{eqnarray}
&&(2f_{xxx}ff_{xx}+f_{xxxxx}f^2-3f^2f_{xyy}-6f_{xx}^2f_x
-5f_{xxxx}ff_x+3ff_xf_{yy}+6ff_{y}f_{xy}+8f_{xxx}f_x^2-6f_xf_{y}^2)a\nonumber\\
&& \quad
+(2f_{xx}f_{xxy}f-f_{yyy}f^2-2f_{xx}^2f_{y}-4f_{xx}f_{xy}f_x+f_{xxxxy}f^2-4f_{xxxy}ff_x
-f_{xxxx}f_{y}f+3f_{yy}f_{y}f
\nonumber\\
&& \quad
+4f_{xxy}f_x^2+4f_{xxx}f_{y}f_x-2f_{y}^3)b
-f_{xxt}f^2+2f_{xt}f_xf+f_{t}f_{xx}f-2f_{t}f_x^2=0. \label{Trif}
\end{eqnarray}
The trilinear equation \eqref{Trif} can not be directly bilinearized. However, if we introduce an auxiliary variable $\tau$ such that
\begin{eqnarray}
[D_xD_{\tau} +a(3 D_y^2-D_x^4)]f\cdot f=0, \label{BKP}
\end{eqnarray}
then the trilinear equation \eqref{Trif} can be rewritten as,
\begin{equation*}
(f\partial_x-2f_x)[a(2bD_x^3D_y-3D_xD_t+3D_xD_{\tau})+bD_yD_{\tau}]f\cdot f=0,
\end{equation*}
which can be solved by the bilinearized equation
\begin{equation}
[a(2bD_x^3D_y-3D_xD_t+3D_xD_{\tau})+bD_yD_{\tau}]f\cdot f=0. \label{JM}
\end{equation}
In \eqref{BKP}--\eqref{JM}, the Hirota's bilinear operator $D_x$ is defined by
\begin{eqnarray}
D_x^n f\cdot g= \left.(\partial_x-\partial_{x'})^n f(x)g(x')\right|_{x'=x}, \label{Dx}
\end{eqnarray}
and the other operators $\{D_y,\ D_t,\ D_{\tau}\}$ are defined in the same way.

Finally, the multiple soliton solutions of the cKP3-4 equation \eqref{KP34} are given by
\eqref{ru} with $f$ being a solution of the trilinear equation \eqref{Trif} which can be solved by the bilinear equation system \eqref{BKP} and \eqref{JM}. The procedure solving the bilinear equation system \eqref{BKP} and \eqref{JM} is standard and well known. We just write down the final result in a fully space-time reversal symmetric form \cite{SF1,SF2},
\begin{eqnarray}
&&f=\sum_{\{\nu\}}K_{\{\nu\}}\cosh\left(\frac12\sum_{i=1}^N\nu_i\xi_i\right),\label{Soliton}\\
&&\xi_i=k_ix+l_iy+\frac{a}{k_i}(k_i^4-3l_i^2)(t+\tau)+\frac{bl_i}{k_i^2}(k_i^4-l_i^2)t+\xi_{i0}, \label{xi}\\
&&K_{\{\nu\}}=\prod_{i<j}a_{ij},\ a_{ij}\equiv \sqrt{k_i^2k_j^2(k_i-\nu_i\nu_jk_j)^2+(k_il_j-k_jl_i)^2},\label{Knu}
\end{eqnarray}
where the summation on $\{\nu\}\equiv \{\nu_1,\ \nu_2,\ \ldots,\ \nu_N\}$ must be done for all non-dual permutations $\nu_i=1,\ -1, \ i=1,\ 2\ \ldots,\ N$. Because the cosh function used in \eqref{Soliton} is an even function, we call two permutations, $\{\nu\}$ and $-\{\nu\}$, are dual each other. The parameters $k_i,\ l_i$ and $\xi_{i0}$ in \eqref{xi} are arbitrary constants. In the left part of this paper, the auxiliary parameter $\tau$ in \eqref{xi} will be taken as zero because it can be absorbed by the arbitrary parameters $\xi_{i0}$.

Especially, one-soliton solution of \eqref{KP34} can be written as
\begin{equation}
u=\frac12 k^2 \mbox{\rm sech}^2\left(\frac{\xi}2\right),\ \xi=kx+ly+\frac{a}{k}(k^4-3l^2)t+\frac{bl}{k^2}(k^4-l^2)t+\xi_{0}\label{1s}
\end{equation}
with arbitrary constants $k,\ l$ and $\xi_0$.

Two-soliton solution possesses the form
\begin{equation}
u=\frac{a_{12}^{+}a_{12}^{-}\left[k_2^2\cosh({\xi_1})+k_1^2\cosh({\xi_2})\right]
+k_2^2k_1^2(k_1^2-k_2^2)^2+(k_1^2+k_2^2)(k_1l_2-k_2l_1)^2}
{2\left[a_{12}^{-}\cosh\left(\frac{\xi_1+\xi_2}2\right)
+a_{12}^{+}\cosh\left(\frac{\xi_1-\xi_2}2\right)\right]^2}\ \label{2s}
\end{equation}
with $\xi_i,\ i=1,\ 2$ being given by \eqref{xi} and
\begin{equation}
a_{ij}^{\pm}=\sqrt{k_i^2k_j^2(k_i \pm k_j)^2+(k_il_j-k_jl_i)^2}. \label{a12}
\end{equation}

\section{Soliton Molecules and travelling wave solutions of \eqref{KP34}}
To find possible soliton molecule solutions, one can use the velocity resonant mechanism \cite{LouS1,LouS2}. For the (2+1)-dimensional KdV equation \eqref{KP34} the soliton resonant conditions read
\begin{equation}
\frac{k_i}{k_j}=\frac{l_i}{l_j}=\frac{\frac{a}{k_i}(k_i^4-3l_i^2)+\frac{bl_i}{k_i^2}(k_i^4-l_i^2)}
{\frac{a}{k_j}(k_j^4-3l_j^2)+\frac{bl_j}{k_j^2}(k_j^4-l_j^2)},\ k_i\neq \pm k_j. \label{vr}
\end{equation}
The solution of \eqref{vr} possesses the form
\begin{equation}
l_i=-\frac{ak_i}{b},\ l_j=-\frac{ak_j}{b}. \label{lij}
\end{equation}
Under the resonance condition \eqref{lij}, two-soliton solution \eqref{2s} becomes a two-soliton molecule
\begin{equation}
u=\frac{(k_1^2-k_2^2)\left[k_2^2\cosh({\xi_1})+k_1^2\cosh({\xi_2})+k_1^2-k_2^2\right]}
{2\left[(k_1-k_2)\cosh\left(\frac{\xi_1+\xi_2}2\right)
+(k_1+k_2)\cosh\left(\frac{\xi_1-\xi_2}2\right)\right]^2}\ \label{1sm}
\end{equation}
with
\begin{equation}
\xi_i=k_i\left(x-\frac{a}by-2\frac{a^3}{b^2}t+\xi_{i0} \right),\ i=1,\ 2.\label{xi12}
\end{equation}
From the resonance condition \eqref{lij} and the travelling variable \eqref{xi12}, one can find some interesting things.

The first one is that the resonance condition \eqref{lij} shows us that this kind of soliton molecules are valid only for the cKP3-4 equation \eqref{KP34} but not for both the KP3 and the KP4 equations.

The second interesting result is that the resonance can be happened for any numbers of solitons with the resonant conditions
\begin{equation}
l_{i_j}=-\frac{ak_{i_j}}{b},\ j=1,\ 2,\ \ldots n, \label{lijn}
\end{equation}
and travelling wave variables
\begin{equation}
\xi_{i_j}=k_{i_j}\left(x-\frac{a}by-2\frac{a^3}{b^2}t+\xi_{i_j0} \right),\ j=1,\ 2,\ \ldots,\ n.\label{xi12a}
\end{equation}
Fig. 1 displays the evolution of a special four-soliton molecule for the field $u$ expressed by \eqref{ru} with $f$ being given by \eqref{Soliton} for $N=4$, the resonant conditions \eqref{lijn} for $n=4$ and the parameter selections
\begin{equation}
a=b=1,\ k_1=\frac{7}{16},\ k_2=\frac38,\ k_3=\frac14,\ k_4=\frac{7}{40},\ \xi_{10}=0,\ \xi_{20}=\xi_{40}=-15,\ \xi_{30}=12.\label{para1}
\end{equation}
\input epsf
\begin{figure}[htbp]
\centering
{\includegraphics[height=4.5cm,width=5.5cm]{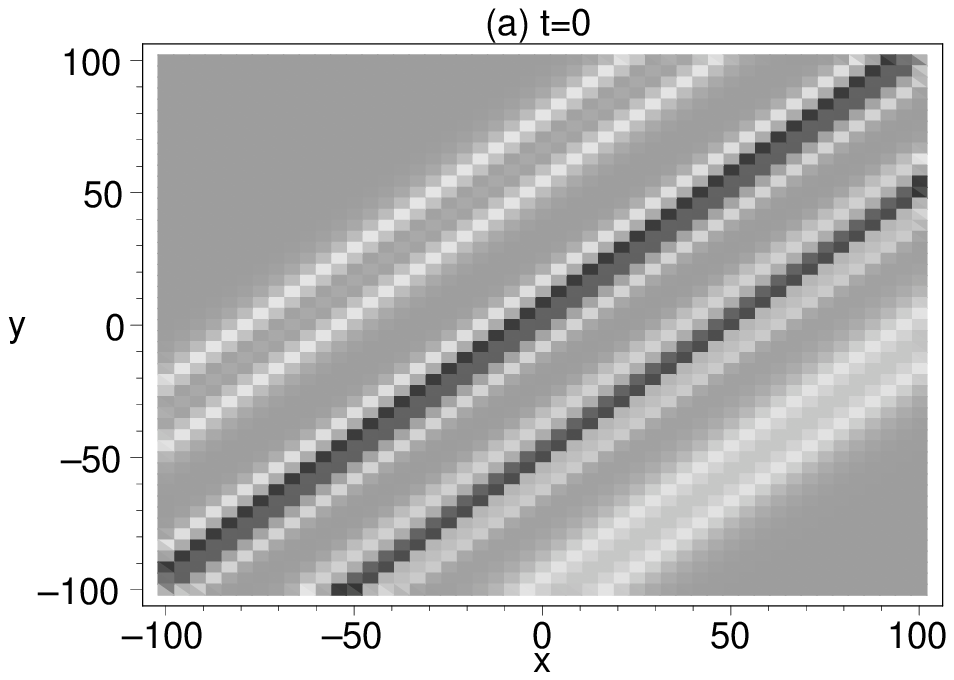}}
{\includegraphics[height=4.5cm,width=5.5cm]{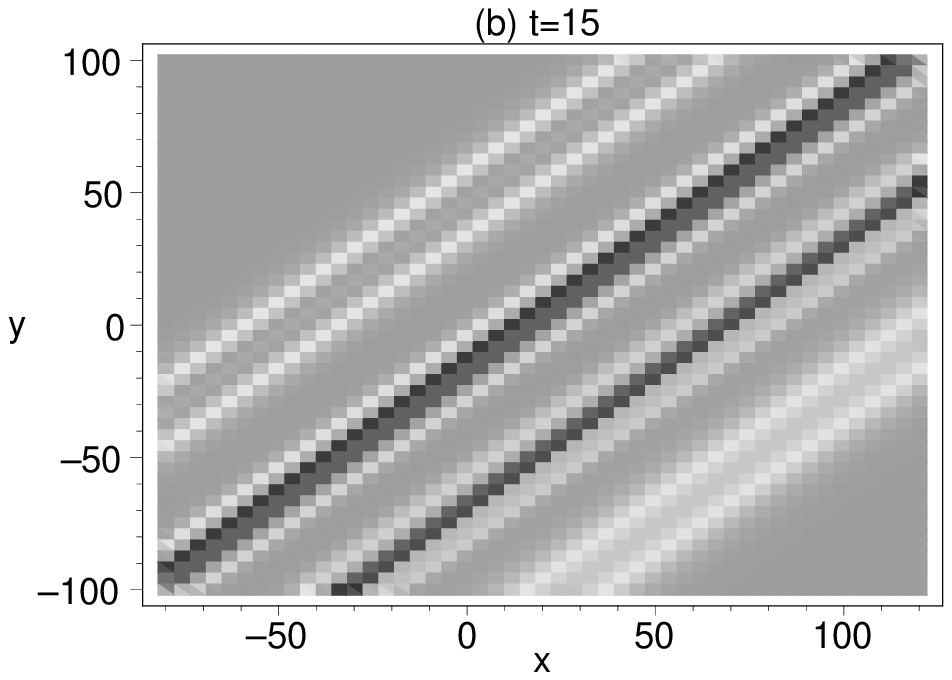}}
\caption{(a) Four-soliton molecue expressed by \eqref{ru} with \eqref{Soliton}, \eqref{lijn}, \eqref{para1} and $N=n=4$ at time $t=0$. (b) Same as in (a) but at $t=15$.}\label{F1}
\end{figure} 
Fig. 2 shows the interaction between one soliton and one three-soliton molecule for the field $u$ expressed by \eqref{ru} with $f$ being given by \eqref{Soliton} for $N=4$, the resonant conditions \eqref{lijn} for $n=3$ and the parameter selections
\begin{equation}
a=b=l_4=1,\ k_1=\frac{7}{16},\ k_2=\frac38,\ k_3=\frac14,\ k_4=\frac{7}{40},\ \xi_{10}=0,\ \xi_{20}=\xi_{40}=-15,\ \xi_{30}=12.\label{para2}
\end{equation}
\input epsf
\begin{figure}[htbp]
\centering
{\includegraphics[height=4.5cm,width=5.5cm]{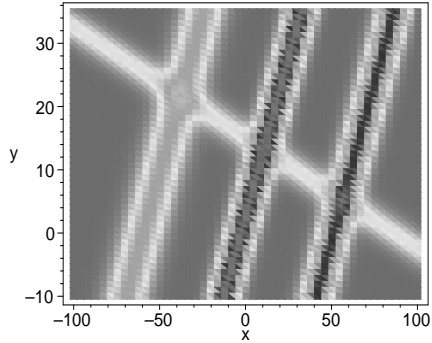}}
\caption{The interactiona between one soliton and one three-soliton molecule for the field $u$ given by \eqref{ru} with \eqref{Soliton}, \eqref{lijn}, \eqref{para2}, $N=4$ and $n=3$ at time $t=0$.}\label{F2}
\end{figure}

The third interesting fact is that the arbitrariness of the resonance for the numbers of solitons hints us the (2+1)-dimensional KdV equation \eqref{KP34} may possess more abundant soliton (and soliton molecule) structures. To check this possibility, we study the travelling wave solutions of the cKP3-4 equation \eqref{KP34} in the form
\begin{equation}
u=U(kx+ly+\omega t)=U(\xi). \label{U}
\end{equation}
Substituting \eqref{U} into \eqref{KP34} yields
\begin{equation}
(\omega k^2+b l^3+3akl^2) U_{\xi}-k^2(ak+bl)(k^2U_{\xi\xi\xi}+6UU_{\xi})=0. \label{Uxi}
\end{equation}
From Eq. \eqref{U}, one can directly find that if the parameters $l$ and $\omega$ are taken as
\begin{equation}
l=-\frac{a}{b}k,\ \omega=-2\frac{a^3}{b^2}k, \label{lo}
\end{equation}
then $U(\xi)$ becomes an arbitrary function. In other words, similar to the linear wave motion equation \eqref{WME} the arbitrary travelling wave moving to a special direction (vertical to $ay-bx=0$ for \eqref{KP34}) in a fixed velocity $c$ ($c=2a^2\sqrt{a^2+b^2}/b^2$ for \eqref{KP34}) may exist in nonlinear cases. Thus, the question proposed in the introduction section possesses a positive answer, the missing D'Alembert type solutions can be re-found in nonlinear systems, say, the cKP3-4 equation.

Eq. \eqref{lo} is just the velocity resonant condition of the soliton molecules. Thus, the soliton molecules obtained form the multiple solitons with the velocity resonant mechanism are just the special case of the arbitrary travelling wave in the form
\begin{equation}
u=u_1\left(x-\frac{a}by-\frac{2a^3}{b^2}t\right).\label{U1}
\end{equation}
It should be emphasized that this kind of arbitrary travelling wave exists only for the cKP3-4 equation \eqref{KP34} but not KP3 equation ($b=0$) and the KP4 equation ($a=0$).

Fig. 3 exhibits some special examples of \eqref{U1} with $a=b=1$.\\
\em Example 1. Few cycle solitons and envelope solitons. \rm
\begin{equation}
u_{ex_1}=\sech(\eta)^2\cos(c\eta),\ \eta\equiv 2t-x+y.\label{ex1}
\end{equation}
The solution \eqref{ex1} displays the few cycle soliton structure for small $c$ (Fig.3a for $c=4$) and the envelope soliton structure for large $c$ (Fig.3b for $c=50$). Few cycle solitons have been found in nonlinear optic systems both in experiments \cite{SPE} and in theories \cite{SunYY,LinJ,LinJ1} but have not yet been reported in the KP hierarchy. Envelope solitons usually appeared in complex systems like the nonlinear Schr\"odinger equation but not in real systems. \\
\em Example 2. Kink solitons and periodic kink (PK) solitons. \rm
\begin{equation}
u_{ex_2}=\tanh(\eta)[1-c\cos(\eta)].\label{ex2}
\end{equation}
The expression \eqref{ex2} is a kink soliton for $c=0$ and a periodic kink for nonzero $c$ (Fig.3c for $c=0.05$). \\
\em Example 3. Kink-Kink molecules and PK-HPK molecules. \rm
\begin{equation}
u_{ex_3}=-\frac{k_1 \exp(k_1\eta)+k_2\exp[k_2(\eta+x_0)]}{1+\exp(k_1\eta)+\exp[k_2(\eta+x_0)]}[1-c\cos(k\eta)].\label{ex3}
\end{equation}
A half periodic kink (HPK) is defined as a kink possessing the property such that it tends to a constant on one side of the kink center and tends to a periodic wave on the other side of the kink center \cite{YZW}.
The solution \eqref{ex3} expresses the kink-kink molecules for $\{c=0,\ k_2\neq k_1\}$ (as shown in Fig.3d for $c=0,\ k_1=-1,\ k_2=-2$ and $x_0=8$) and the PK-HPK molecules (as shown in Fig.3e for $c=0.06,\ k_1=-1,\ k_2=-2, k=5$ and $x_0=10$). \\
\em Example 4. Dissipative solitons, kink-antikink molecules and PK-PAK (periodic kink and periodic antikink) molecules. \rm
\begin{equation}
u_{ex_4}=\frac12+\frac12[c_1\cos(c\eta)-1]\tanh(\eta+y_0)\tanh(\eta-x_0).\label{ex4}
\end{equation}
The expression \eqref{ex4} displays the dissipative soliton (or kink-antikink molecule) structure for $c_1=0$ or $c=0$) and the PK-PAK molecule (or periodic dissipative soliton) structure for $cc_1\neq 0$. Fig. 4f is a plot of a special PK-PAK molecule expressed by \eqref{ex4} with the parameter selections
$$x_0=-5,\ y_0=-5,\ c_1=0.1,\ c=5$$
at time $t=0$.
\input epsf
\begin{figure}[htbp]
\centering
{\includegraphics[height=4.5cm,width=5.5cm]{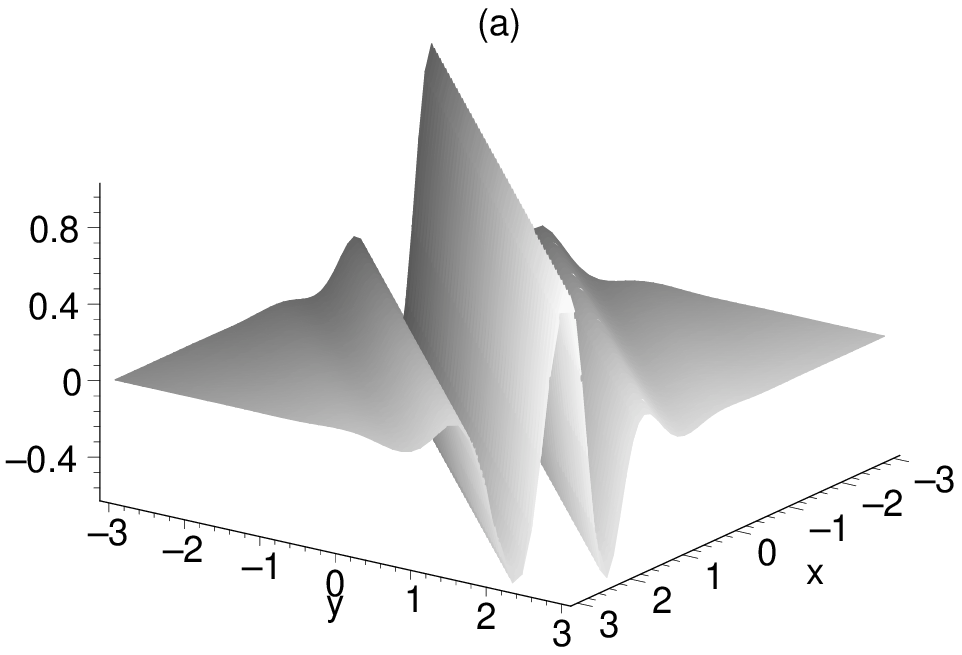}}
{\includegraphics[height=4.5cm,width=5.5cm]{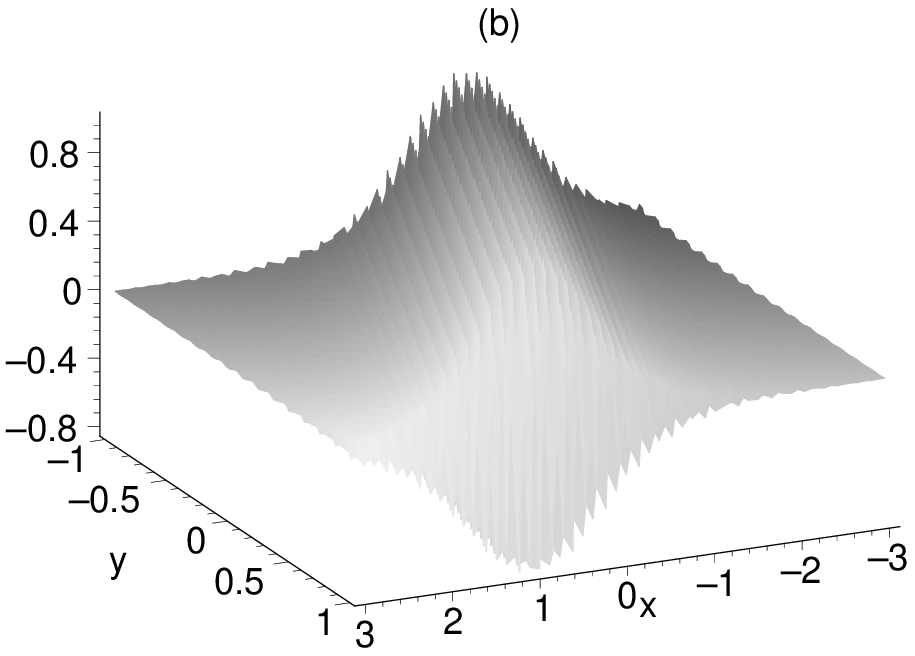}}
{\includegraphics[height=4.5cm,width=5.5cm]{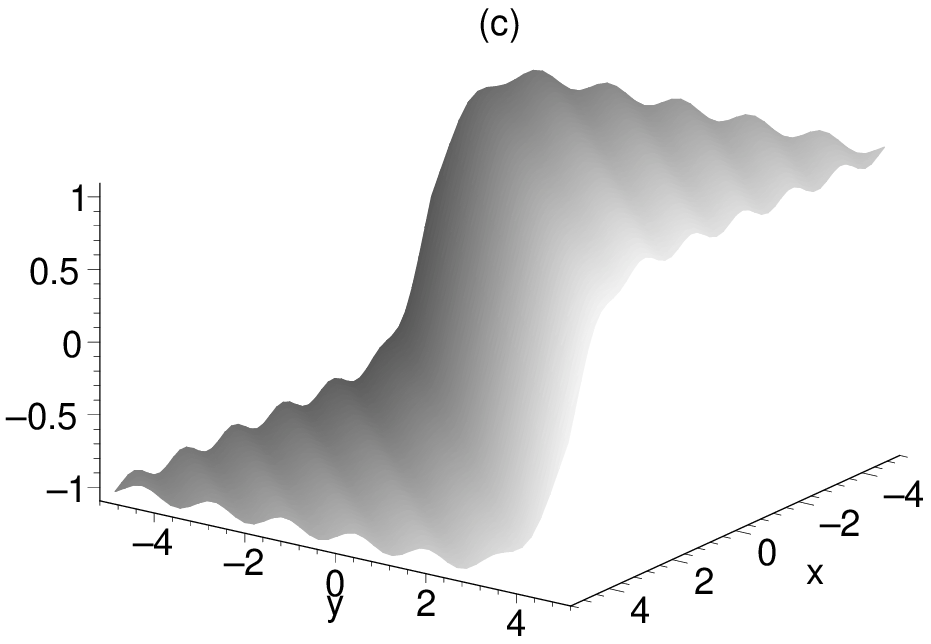}}
{\includegraphics[height=4.5cm,width=5.5cm]{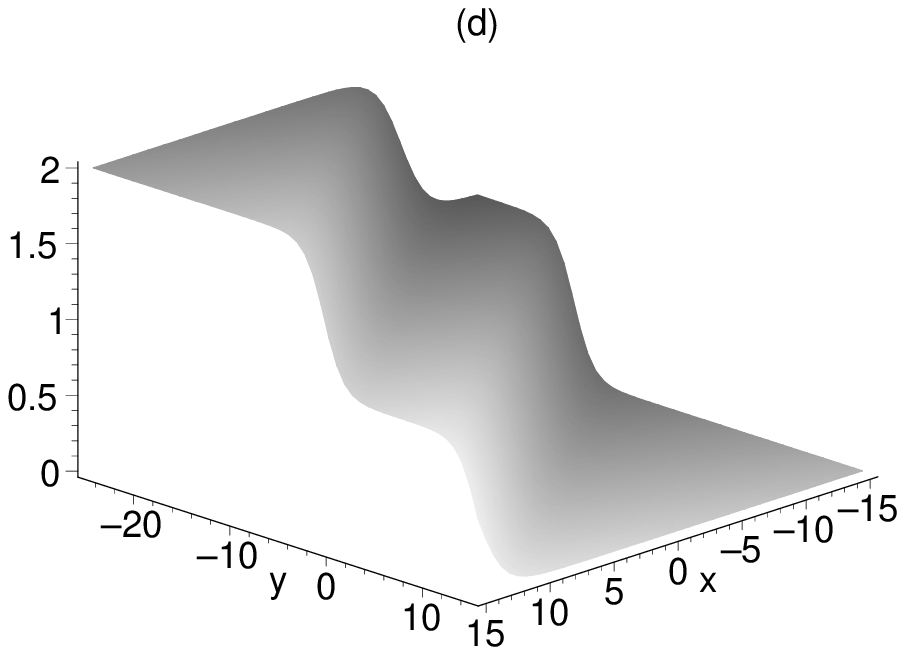}}
{\includegraphics[height=4.5cm,width=5.5cm]{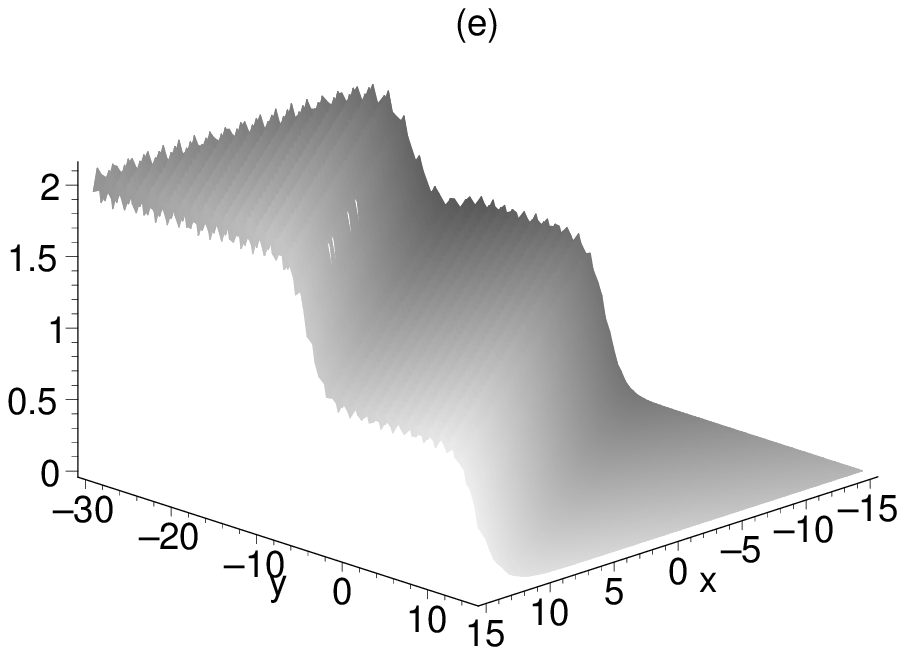}}
{\includegraphics[height=4.5cm,width=5.5cm]{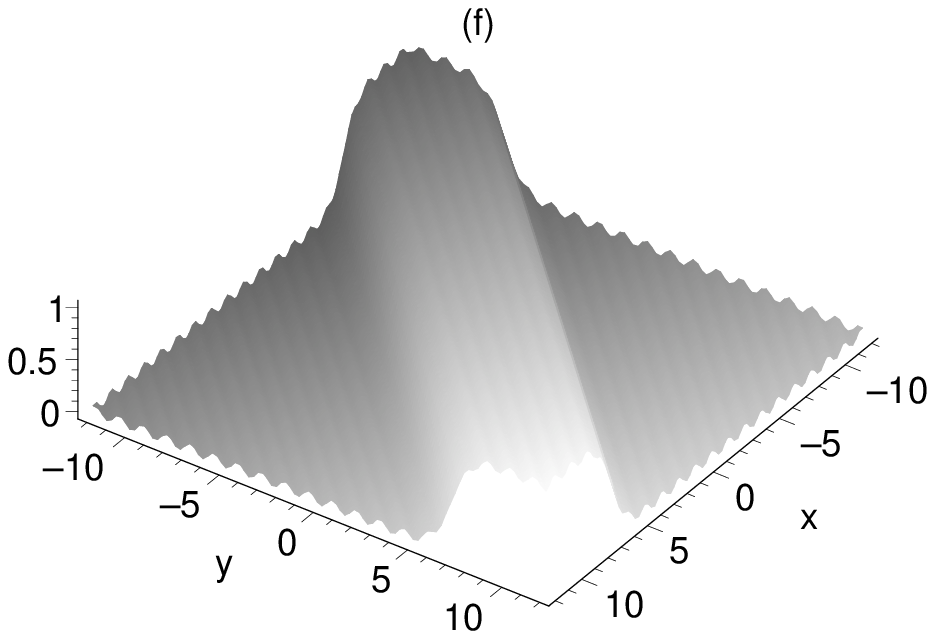}}
\caption{(a) Few cycle soliton structure expressed by \eqref{ex1} with $c=4$ at time $t=0$. (b) Few cycle soliton structure expressed by \eqref{ex1} with $c=50$ at time $t=0$. (c) Periodic kink given by \eqref{ex2} with $c=0.5$ at time $t=0$. (d) Kink-kink molecule given by \eqref{ex3} with $c=0,\ k_1=-1,\ k_2=-2$ and $x_0=8$. (e) Kink-kink molecule given by \eqref{ex3} with $c=0.06,\ k_1=-1,\ k_2=-2,\ k=5$ and $x_0=10$ at $t=0$. (f) PK-APK molecule given by \eqref{ex4} with $x_0=y_0=-c=-5$ and  $c_1=0.1$ at time $t=0$.}\label{F3}
\end{figure}

Without the condition \eqref{lo}, the solution of \eqref{Uxi} can be written as
\begin{equation}
U(\xi)=\frac{3 a k l^2+b l^3+k^2\omega}{6k^2(ak+bl)}-\frac23 c^2 k^2 (2 m^2-1) +2 c^2 k^2 m^2 \mbox{\rm cn}^2(c\xi-\xi_0,\ m)\label{cn}
\end{equation}
with arbitrary constants $c,\ \xi_0,\ m,\ k,\ l$ and $\omega$. In additional to the arbitrary travelling wave solution \eqref{U1}, the general travelling wave of the cKP3-4 \eqref{cn} is a periodic wave for $m\neq 1$ and/or a soliton solution for $m=1$.

\section{Conclusions and discussions\vspace{-0.3em}}
In summary, there are some different types of (2+1)-dimensional KdV extensions. In this paper, it is found that in addition to the KP3 equation, the KP4 equation is also an extension of the KdV equation in (2+1) dimensions. Thus, the combination of the KP3 equation and the KP4 equation (i.e. the cKP3-4 equation \eqref{KP34}) is also a (2+1)-dimensional KdV extension. The Lax pair and the dual Lax pair of the cKP3-4 equation are explicitly given.  The cKP3-4 equation can be bilinearized by introducing an auxiliary parameter. The multiple soliton solutions can be directly written down with help of the bilinearized equation system and the auxiliary parameter can be absorbed by the soliton position parameters.

The cKP3-4 equation is quite different from any member of the KP hierarchy, say, the KP3 and KP4 equations of the hierarchy \eqref{2+1}. Some interesting properties are valid only for the combined system \eqref{KP34} but not for the separated KP3 and KP4 systems. For instance, the soliton molecules exist only for the cKP3-4 but not for the KP3 and KP4 equations. Any numbers of solitons can be involved in one soliton molecule. 
The D'Alembert type solutions (arbitrary travelling wave moving in one direction with fixed model dependent velocity) including various new types of solitons and soliton molecules can be found for the cKP3-4 equation. This kind of arbitrariness exists for linear waves but it has not yet found for any other known nonlinear systems. The more about the cKP3-4 equation \eqref{KP34}, say the interactions among the special local excitations like shown in Fig. 3 and the usual solitons \eqref{ru} with \eqref{Soliton}, should be further studied.

\section*{Acknowledgements\vspace{-0.2em}}
The work was sponsored by the National Natural Science Foundations of China (Nos.11975131,11435005) and K. C. Wong Magna Fund in Ningbo University.


\vspace{-1em}
\begin{thebibliography}{00}
\setlength{\itemsep}{-0.75ex}
\vspace{-0.5em}
\bibitem{KdV}D. G. Crighton,
Appl. Math. \bf 39, \rm 39 (1995).
\bibitem{Guo} H. Y. Guo, Z. H. Wang and K. Wu,
Phys. Lett. B \bf 264, \rm 277 (1991).
\bibitem{IST} C. S. Gardner, J. M. Greene, M. D. Kruskal and R. M. Miura,
Phys. Rev. Lett. \bf 19, \rm 1095 (1967).
\bibitem{Hirota}
R. Hirota, The direct method in soliton theory, Edited and translated by A. Nagai, J. Nimmo, C. Gilson, Cambridge Tracts
in Mathematics No. 155, Edition 1 (Cambrifge: Cambridge University Press) pp.1-61 (2004).
\bibitem{Gu} C. H. Gu, H. S. Hu and Z. X. Zhou, Darboux Transformations in Integrable Systems: Theory and their Applications to Geommetry,
Edition 1 (Dordrecht, Netherland: Springer) pp 1-64 (2005).
\bibitem{KP}B. B. Kadomtsev and V. I. Petviashvili, Sov. Phys. Dokl. \bf 15,\ \rm 539 (1970).
\bibitem{NNV1} L.P. Nizhnik, Sov. Phys. Dokl. \bf 25, \ \rm 706 (1980).
\bibitem{NNV2} A.P. Veselov, S.P. Novikov, Sov. Math. Dokl. \bf 30, \rm 588 (1984).
\bibitem{NNV3} S.P. Novikov, A.P. Veselov, Physica D \bf 18, \rm 267 (1986).
\bibitem{ANNV1} S. Y. Lou and X. B. Hu, J. Math. Phys. \bf 38,\ \rm 6401 (1997).
\bibitem{ANNV2} S. Y. Lou and H-y Ruan,
J. Phys. A: Math. Gen. \bf 35,\ \rm 305 (2001).
\bibitem{ANNV3} M. Boiti, J. J. P. Leon , M. Manna and F. Pempinelli, Inverse Problems \bf 2, \rm 271 (1986).
\bibitem{Ito}M. Ito, J. Phys. Soc. Jpn. \bf 49, \rm 771 (1980).
\bibitem{BS}G. I. Bogoyavlenskii, Russ. Math. Surveys, \bf 45,\ \rm 1 (1990).
\bibitem{BS1}S. Y. Lou, Commun. Theor. Phys. \bf 28, \rm 41 (1997).
\bibitem{SP05}M. Stratmann, T. Pagel and F. Mitschke, 
Phys. Rev. Lett. \bf 95,\ 143902 (2005). 
\bibitem{HK17}G. Herink, F. Kurtz, B. Jalali, D. R. Solli and C. Ropers,
Science  \bf 356,\ \rm 50 (2017).
\bibitem{LXM}X. M. Liu, X. K. Yao and Y. D. Cui,
Phys. Rev. Lett. \bf 121,\ \rm 023905 (2018).
\bibitem{Nano} C. Wang, L. Wang, et al,
Nanotechnol. \bf 30,\ \rm 025204 (2019).
\bibitem{LNS} K. Lakomy, R. Nath and L. Santos,
Phys. Rev. A \bf 86, \rm 013610 (2012).
\bibitem{SA} J. S. Peng, S. Boscolo, Z. H. Zhao, H. P. Zeng,
Sci. Adv. \bf 5, \rm 1110 (2019).
\bibitem{CK03} L. C. Crasovan, Y. V. Kartashov,  D. Mihalache, L. Torner, Y. S. Kivshar, V. M. Perez-Garcia,
Phys. Rev. E \bf 67,\ 046610 (2003).
\bibitem{YB11} C. Yin, N. G. Berloff, V. M. Perez-Garcia, D. Novoa, A. V. Carpentier, H. Michinel,
Phys. Rev. A \bf 83,\ \rm 051605 (2011).
\bibitem{LouS1}S. Y. Lou, Soliton molecules and asymmetric solitons in fuid systems via velocity resonance, arXiv: 1909.03399. (2019).
\bibitem{LouS2}D. H. Xu and S. Y. Lou, Acta Phys. Sin. \bf 69,\ \rm 014208 (2020) (in Chinese).
\bibitem{KdV5}A. S. Fokas and Q. M. Liu, Phys. Rev. Lett. \bf 77,\ \rm 2347 (1996).
\bibitem{LiB}Z. Zhang, X. Y. Yang and B. Li, Appl. Math. Lett. \bf 103, \rm 106168 (2020).
\bibitem{XYT}C. J. Cui, X. Y. Tang and Y. J. Cui, Appl. Math. Lett. 103 (2020) 106109.
\bibitem{YZW}Z. W. Yan and S. Y. Lou, Soliton molecules in Sharma-Tasso-Olver-Burgers equation arXiv: 1912. 13324.nlin.PS (2019).
\bibitem{PRLou}S. Y. Lou, Phys. Rev. Lett. \bf 71,\ \rm 4099 (1993).
\bibitem{JPA93}S. Y. Lou,  J. Phys. A: Math. Gen. \bf 26,\ \rm 4387 (1993).
\bibitem{JMP94}S. Y. Lou, J. Math. Phys. \bf 35,\ \rm 1755 (1994).
\bibitem{NKP1}S. Y. Lou,
Physica Scripta \bf 57,\ \rm 481 (1998).
\bibitem{NKP2}X. B. Hu, S. Y. Lou and X. M. Qian,
Stud. Appl. Math. \bf 122,\ \rm 305 (2009).
\bibitem{SF1}S. Y. Lou, J. Math. Phys. \bf 59,\ \rm 083507 (2018).
\bibitem{SF2}S. Y. Lou, Acta Phys. Sin. \bf 69,\ \rm 010503 (2020) (in Chinese).
\bibitem{SPE}
S. J. Im, A. Husakou and J. Herrmann
Phys. Rev. A \bf 82,\ \rm 025801 (2010).
\bibitem{SunYY}Y. Y. Sun and H. Wu, Phys. Scr. \bf 88,\ \rm 065001 (2013).
\bibitem{LinJ}Z. J. Gao and J. Lin,
   Opt. Express, \bf 26,\ \rm 9027 (2018).
\bibitem{LinJ1}Z. J. Gao, H. J. Li and J. Lin,
J. Opt. Soc. Am. B \bf 36,\ \rm 312 (2019).
\end{thebibliography}
\end{document}